\documentstyle[preprint,aps]{revtex}

\begin{document}

\preprint{\tighten
\vbox{
\hbox{FSUHEP-960315}
\hbox{UR-1457}
\hbox{UH-511-845-96}
\hbox{March 1996}
}} 
\title{Simultaneous Search for Two Higgs Bosons \\
of Minimal Supersymmetry at the LHC}
\author{Salavat Abdullin$^a$, Howard Baer$^b$, Chung Kao$^c$, 
Nikita Stepanov$^a$ and Xerxes Tata$^d$}
\address{
$^a$Institute for Theoretical and Experimental Physics, 
Moscow, Russia \\
$^b$Department of Physics, Florida State University, 
Tallahassee FL 32306, USA \\
$^c$Department of Physics and Astronomy, University of Rochester, 
Rochester NY 14627, USA \\
$^d$Department of Physics and Astronomy, University of Hawaii, 
Honolulu HI 96822, USA}
\maketitle
\begin{abstract}

The prospects of detecting the CP-odd Higgs pseudoscalar ($A$) 
in the minimal supersymmetric model 
via its decay into a $Z$ boson and the lighter CP-even Higgs scalar ($h$)
at the CERN Large Hadron Collider are investigated.
The final state of $Z \to l^+l^-$ and $h \to b\bar{b}$,
may provide a promising way to simultaneously detect the $A$ and the $h$. 
The compact muon solenoid detector performance is adopted 
for a realistic study of observability.
In this discovery channel, the masses of the $h$ and the $A$ 
can be reconstructed.
The impact of supersymmetric decay modes is considered.

\end{abstract}


\section{Introduction}

The most important experimental goal of the CERN Large Hadron Collider (LHC) 
is to unravel the mechanism of electroweak symmetry breaking. 
In the Standard Model (SM) of electroweak interactions, 
only one Higgs doublet is required to generate mass for fermions 
as well as gauge bosons. One neutral CP-even Higgs boson ($H^0$) 
appears after spontaneous symmetry breaking. 
Various extensions of the SM have more complicated Higgs sectors 
and lead to additional physical spin zero fields \cite{GUIDE}.

The minimal supersymmetric extension of the Standard Model (MSSM) \cite{MSSM}
has two Higgs doublets with vacuum expectation values $v_1$ and $v_2$.
After spontaneous symmetry breaking, there remain five physical Higgs bosons:
a pair of singly charged Higgs bosons $H^{\pm}$,
two neutral CP-even scalars $H$ (heavier) and $h$ (lighter),
and a neutral CP-odd pseudoscalar $A$.
The Higgs sector is strongly constrained by supersymmetry
so that, at the tree level, all Higgs boson masses and couplings are
determined by just two independent parameters,
which are commonly chosen to be the mass of the CP-odd pseudoscalar ($M_A$)
and $\tan \beta \equiv v_2/v_1$.

At the one loop level, 
radiative corrections from the $t$-quark and the $b$ quark Yukawa couplings 
substantially modify the tree level formulae for masses 
and mixing patterns in the Higgs sector.
Let us briefly review recent studies on the search for the $A$ at the LHC.
In most studies, the parameters were selected such that 
supersymmetric particle (SUSY particle) masses were large 
so that Higgs boson decays to SUSY particles 
were kinematically forbidden. 
The $A \to \gamma\gamma$ mode might be observable at the LHC 
for 120 GeV $< M_A \le 2m_t$ and $\tan\beta$ close to one 
\cite{HGG}-\cite{KZ}.
The $A$ does not have tree level couplings with gauge boson pairs.
The one-loop induced $A \to ZZ \to l^+l^-l^+l^-$ decay mode 
was found to be a useful discovery channel at the LHC only if 
$\tan\beta$ is much less than one and $M_A$ is near the $t\bar t$ threshold 
\cite{AZZ}.
If $\tan\beta$ is large, the $\tau\bar{\tau}$ decay mode 
might be a promising discovery channel for the $A$ in the MSSM 
\cite{KZ,CMS,ATLAS}.
It has been suggested that the $A$ might also be observable via
its $b\bar{b}$ decays in a large region of the ($M_A,\tan\beta$) plane,
provided that sufficient $b$-tagging capability can be achieved \cite{DGV,Hbb}.
For $\tan\beta$ larger than about 10, 
the $A \to \mu^+\mu^-$ discovery mode at the LHC was found 
to be observable in a large region of the $(M_A,\tan\beta)$ parameter space 
\cite{CMS,mumu}, 
even when Higgs decays to SUSY particles are significant \cite{mumu}.
This channel will provide a good opportunity to reconstruct the masses 
for neutral Higgs bosons with high precision.

There are, however, regions of parameter space where rates 
for Higgs boson decays to SUSY particles, e.g. the charginos and neutralinos, 
are substantial or even dominant.
While these decays reduce the rates for SM signatures, making
conventional detection of Higgs bosons even more difficult,
they also open up a number of new promising modes for Higgs detection 
\cite{Z2Z2}.
A recent study suggests that if the sleptons are light,  
it  might be possible to search for the neutral Higgs bosons 
via their decays into slepton pairs 
\cite{slepton}.

The MSSM Higgs pseudoscalar ($A$) decays into a $Z$ boson 
and the lightest Higgs scalar ($h$) with a large branching ratio 
for $\tan\beta < 4$ and $M_Z +M_h < M_A \le 2m_t$.\footnote{
The $H \to ZA$ decay is possible for $M_A$ less than about 60 GeV, 
that is slightly above the excluded region of the LEP and 
within the discovery region for the LEP II.}
The final state where $Z \to l^+l^-$ and $h \to \tau^+\tau^-$, 
was found to be a promising channel to simultaneously detect the $A$ 
and the $h$ at the LHC \cite{lltt}.
The $h \to b\bar{b}$ has a branching fraction 
about 20 times larger than that of $h \to \tau\bar{\tau}$.
It is, therefore, natural to ask whether with a good b-tagging efficiency
it is possible to search for the $A$ and the $h$ in the decay of $A \to Zh$ 
with the final state of $Z \to l^+l^-$ and $h \to b\bar{b}$.

In this paper, 
we assess the prospects for the simultaneous discovery of the $A$ 
and the $h$ at the LHC via the process $A \to Zh \to l^+l^- b\bar{b}$.
The compact muon solenoid detector (CMS) performance is adopted 
for a realistic study of observability.\footnote{Recently, similar results 
for this discovery channel were also found 
for the ATLAS detector \cite{ATLAS1}.}
A brief summary of our preliminary analysis, 
for the case when Higgs boson decays to SUSY particles 
are kinematically forbidden, 
has appeared in the CMS Technical Proposal \cite{CMS}. 
In this complete analysis, 
we have carried out detailed studies for dominant backgrounds, 
and further, have considered the impact of Higgs decays to SUSY particles.
Parton level calculations are presented in Section II.
Results from more realistic simulations are discussed in Section III. 
We end with a discussion of our results in Section IV.

\section{Parton Level Calculations}

The total cross section for the process 
$pp \to A \to Zh \to l^+l^-b\bar{b} +X$ 
is evaluated from the cross section $\sigma(pp \to A +X)$ 
multiplied with the branching fractions of $A \to Zh$, 
$Z \to l^+l^-$ and $h \to b\bar{b}$.
The parton distribution functions of CTEQ2L \cite{CTEQ} 
are chosen to evaluate the cross section of $pp \to A +X$ 
with $\Lambda_4 = 0.190$ GeV and $Q^2 = M_A^2$.
We take $M_Z = 91.187$ GeV, $\sin^2\theta_W = 0.2319$,
$M_W = M_Z \cos\theta_W$,
$m_b = 4.7$ GeV, and $m_t = 175$ GeV.


Gluon fusion ($gg \to A$), via the top quark and the bottom quark 
triangle loop diagrams, is the major source for the Higgs pseudoscalar  
if $\tan\beta$ is less than about 4.
The amplitude for the process $gg \to A$ 
is a function of the quark mass squared ($m_q^2$), 
$M_A^2$, and $\tan\beta$ \cite{gggA}:
\begin{enumerate}
\item 
For $M_A \le m_t$, the $t$-loop is almost independent of $m_t$, 
the $ggA$ coupling can be either obtained from the low energy theorem 
of the axial anomaly \cite{Adler}-\cite{Bill} 
or from the exact calculation for $gg \to A$ 
at the limit of $m_t^2/M_A^2 \gg 1$.
\item 
At the threshold of $M_A = 2m_t$, 
the imaginary part of the amplitude turns on. 
Therefore, the cross section 
is significantly enhanced when $M_A$ is close to $2m_t$.
\item 
When $m_q^2$ is much less than $M_A^2$, the amplitude is proportional to 
$m_q^2 [ ln^2(m_q^2/M_A^2) -2i\pi\ln(m_q^2/M_A^2) ]$. 
The $t$-loop dominates in a large region of $\tan\beta$.
The cross section is almost proportional to $\cot^2\beta$ for $\tan \beta < 10$.
Only for $\tan\beta$ close to $m_t/m_b$, can the $b$-loop dominate and the
total cross section be enhanced by large $\tan \beta$.
\end{enumerate}

Since the Yukawa coupling of $A b\bar{b}$ is proportional to $\tan\beta$,
the production rate of the $A$ from $b\bar{b} \to A$ is enhanced 
for large values of $\tan\beta$. 
If $\tan\beta$ is larger than about 7, the $A$ is dominantly produced
from $b$-quark fusion ($b\bar{b} \to A$) \cite{Duane}.
We have evaluated the cross section of the $A$ in $pp$ collisions
$\sigma(pp \to A +X)$,  with two dominant subprocesses:
$gg \to A$ and $gg \to A b\bar{b}$.
The cross section of $gg \to A b\bar{b}$ is a good approximation 
to the `exact' cross section of $b\bar{b} \to A$ \cite{Duane} 
for $M_A$ less than about 500 GeV.
QCD radiative corrections which increase the gluon fusion ($gg \to A$) 
production cross section by about 50\% to 80\% for $\tan\beta \sim 1$ 
\cite{Kauffman,Spira} 
are not included in our computation for either the signal or the backgrounds.
 

We have included complete one loop corrections 
from both the top and the bottom Yukawa interactions 
to the Higgs masses and couplings 
using the effective potential \cite{OYY}-\cite{Haber}. 
The formulae presented in Ref. \cite{Mike} have been employed 
in our calculations.
The contributions from the D-terms are usually small \cite{Z2Z2,Mike},
and therefore, are not included.
The soft SUSY breaking parameters $A_t$ and $A_b$ are taken to be zero.
We take $m_{\tilde{g}} \simeq m_{\tilde{l}} \simeq m_{\tilde{q}} = |\mu|$ 
and consider three sets of parameters: 
(a) $\mu= 1000$ GeV, 
such that the  Higgs boson decays to SUSY particles are kinematically forbidden;
(b) $|\mu| = 500$ GeV, 
such that the Higgs boson decays to SUSY particles become significant; and 
(c) $|\mu| = 300$ GeV, 
such that the Higgs boson decays to SUSY particles are large and dominant 
when $\tan\beta$ is less than about 10.
 
At first, 
let us take 
$m_{\tilde{g}} \simeq m_{\tilde{l}} \simeq m_{\tilde{q}} = \mu = 1000$ GeV, 
such that the neutralinos ($\tilde{Z}_i, i = 1-4$), 
the charginos ($\tilde{W}_j, j = 1,2$), 
the squarks ($\tilde{q}$), and the sleptons ($\tilde{l}$), are heavy 
and the Higgs boson decays to SUSY particles are kinematically forbidden.
With QCD radiative corrections \cite{Braaten,Drees}, 
the branching fraction of $A \to b\bar{b}$ is reduced by about a factor of 2. 
The $b\bar{b}$ mode dominates the Higgs pseudoscalar decays
for $\tan\beta$ larger than about 4 and $M_A \le 2m_t$.

For $M_Z +M_h < M_A \le 2m_t$ and $\tan\beta < 4$, 
the branching fraction of $A \to Zh$ is comparable to $B(A \to b\bar{b})$ 
if the SUSY decay modes are kinematically forbidden. 
The decay of $A \to Zh$ can become dominant if $\tan\beta$ is less than 2.
In this region, the lightest Higgs scalar mass 
is modified mainly by the radiative corrections from the top quark 
Yukawa couplings.
The leading radiative corrections to the $M_h$ 
can be expressed in terms of the third generation doublet and singlet 
squark masses ($m_{\tilde{Q}_L}$ and $m_{\tilde{t}_R}$) \cite{HGG},
\begin{equation}
M_h^2 = \frac{1}{2} [ {M_A}^2+{M_Z}^2 +\delta -\xi^{\frac{1}{2}} ]
\end{equation}
where
\begin{equation}
\xi = [({M_A}^2-{M_Z}^2)\cos 2\beta+\delta]^2 
     +\sin^2 2\beta({M_A}^2+{M_Z}^2)^2,
\end{equation}
and
\begin{equation}
\delta = \frac{3g^2{m_t}^4}{16\pi^2{M_W}^2\sin^2\beta} 
         \times \ln [(1+\frac{ m_{\tilde{t}_R}^2 } { m_t^2 } )
                              (1+\frac{ m_{\tilde{Q}_L}^2 } { m_t^2 } ) ].
\end{equation}
We have, of course, included complete one loop corrections 
from both the top and the bottom Yukawa interactions in our analysis 
with the formulae in Ref. \cite{Mike}. 

The total cross section for the process 
$pp \to A \to Zh \to l^+l^- b\bar{b} +X$ is shown in Fig. 1 
as a function of $M_A$ for various values of $\tan\beta$.
We take $m_{\tilde{g}} \simeq m_{\tilde{l}} \simeq m_{\tilde{q}} = |\mu|$ 
and consider three sets of parameters 
similar to those chosen in Ref. \cite{Z2Z2},
(a) $ m_{\tilde{q}} = \mu = 1000$ GeV, 
such that the Higgs boson decays to SUSY particles are kinematically forbidden;
(b) $ m_{\tilde{q}} = \mu = 300 $ GeV, 
and 
(c) $ m_{\tilde{q}} = -\mu = 300$ GeV,
such that the Higgs boson decays to SUSY particles are large and dominant
for $\tan\beta$ less than about 10.
The signal cross section is largest for case (a) where no supersymmetric
decays of $A$ are kinematically accessible. 
For the $|\mu|=300$~GeV cases (b) and (c), 
the branching fraction for SUSY decays to charginos and neutralinos 
are larger for the case of positive $\mu$ since $\tilde{W}_1$
and $\tilde{Z}_{1,2}$ are generally lighter than when $\mu<0$. 
The reduction in the cross section due to the opening of the SUSY modes 
is smaller when $\tan\beta$ is large because the $b$-Yukawa coupling,
and hence the decay rate via the $b\bar{b}$ channel, grows with $\tan\beta$.
We also remark that for $\mu=300$~GeV and $M_A \sim 300$~GeV, and 
$\tan\beta < 2$, the cross section is further reduced because the decay
$h \to \tilde{Z_1} \tilde{Z_1}$ become accessible, 
so that the $B(h\to b\bar{b})$ is reduced by almost 50\%.

In Fig. 2, we present the total cross section for the process 
$pp \to A \to Zh \to l^+l^- b\bar{b} +X$
as a function of $\tan\beta$ for various values of $M_A$.
Other parameters are the same as those in Fig. 1.
Note that the $M_h$ is usually enhanced with a larger $\tan\beta$.
In Fig. 2(a), the cross section for $M_A = 200$ GeV drops sharply 
for $\tan\beta > 4$ because $M_Z +M_h$ becomes larger than $M_A$.
In Figs. 2(b) and 2(c), the cross section of $M_A = 200$ GeV is smaller than
that of $M_A = 300$ GeV for a large $\tan\beta$ 
because the phase space is suppressed 
when $M_A$ is only slightly larger than $M_Z +M_h$. 
 
Fig. 3 shows the contours of total cross section ($\sigma$) 
for $pp \to A \to Zh \to l^+l^- b\bar{b} +X$ 
in the $(M_A,\tan\beta)$ plane for $\sigma = 1, 10$ and 100 fb. 
The SUSY parameters are the same as those in Fig. 1.
Also shown as the dashed line is the mass contour for $M_A = M_Z +M_h$.
If SUSY decays of the $A$ and the $h$ are kinematically forbidden, 
the cross section can be larger than 100 fb, for $M_Z+M_h < M_A \le 2m_t$, 
and $\tan\beta$ less than about 2.

\section{Realistic Simulations}
 
Recent studies on $\phi \to b\bar{b}, \phi = H,h,A$ \cite{Hbb},
have shown that more realistic estimations with a detector response
could remarkably modify the results of the parton level calculations.
To make such a comparison some detector model has to be employed.
The LHC will have two general purpose detectors: the CMS and the ATLAS.
Both detectors will provide precise muon and electron momentum reconstruction
as well as reasonable jet energy resolution.
We have adopted the CMS detector performance parameters \cite{CMS}
to estimate the signal and backgrounds.
Similar results can be expected for the ATLAS detector \cite{ATLAS}.
 
\subsection{Calculation tools}
 
The PYTHIA 5.7/JETSET 7.4 generator \cite{PYTHIA} 
with the CTEQ2L \cite{CTEQ} parton distribution functions is used 
to simulate events at the particle level.
To simulate the signal, we use PYTHIA to generate the subprocess $gg \to A$,
which is the dominant source of the $A$ for low $\tan\beta$.
But only the kinematics, parton showering and hadronization schemes
are used directly from PYTHIA.
The cross section of $ pp \to A \to l\bar{l} b\bar{b} +X$ is
very sensitive to the choice of MSSM parameters.
To be consistent with the parton level calculations,
the results from PYTHIA are rescaled to the same cross section
and the one-loop corrected $M_h$ is incorporated in PYTHIA by hand.
 
The output from PYTHIA/JETSET is processed
with the CMSJET program \cite{CMSJET}
which has been developed for the fast simulations of the
``realistic'' CMS detector response.
The resolution effects are taken into account
with the parametrizations obtained from the
detailed GEANT \cite{GEANT} simulations.
The CMSJET also includes some analysis routines, in particular,
a set of jet reconstruction algorithms.
All events containing $e^+ e^-$ or
$\mu^+ \mu^- $ pairs and at least two jets
are reconstructed and stored for further analysis.
 
At the first stage, the minimal kinematical cuts (I) are applied:
\begin{eqnarray}
P_T^{\mu} & \ge & 10  \; {\rm GeV}, \nonumber \\
P_T^{e}   & \ge & 20  \; {\rm GeV}, \;\; |\eta^{\mu,e}| \le 2.4, \nonumber \\
E_T^{jet} & \ge & 20  \; {\rm GeV}, \;\; |\eta^{jet}| \le 2, \nonumber \\
|M_{ll} -M_Z| & < & 6 \; {\rm GeV}.
\label{eq:CUT1}
\end{eqnarray}
In principle, the $\eta$ coverage of CMS detector for jets is much better,
but for $|\eta^{jet}| > 2$ the $b$-tagging capability rapidly deteriorates.
The lepton pair serves as an event trigger. The cut on $M_{ll}$ is
included to remove the reducible backgrounds from other dilepton sources.
 
\subsection{$b$-tagging}
 
The $b$-tagging capability of any detector is mainly defined by the
tracker system quality. Preliminary CMS results indicate
that the average $b$-tagging efficiency at the level of 40\%
with a mistagging probability of about 2\% can be achieved quite easily
with the microvertex detectors \cite{CMS}.
The algorithm applied combines common impact parameter
and lepton tagging techniques.
These values look quite reasonable
in light of the recent CDF results \cite{CDFb}.
There is some hope that some more sophisticated $b$-tagging
algorithms will allow one to achieve the $b$-tagging efficiency of 60\%
with a purity of 1\% for both the CMS and the ATLAS experiments.
 
Because of the existing uncertainties we prefer to use here
a rather artificial and simplified but more general way to take
into account $b$-tagging.
Namely, all jets are matched as a $b$-jet or a non $b$-jet
when they are generated.
In our analysis, a given jet passing
kinematical cuts is matched as a $b$-jet if there is a $b$-quark in
a cone $\Delta R \le 0.3$,
where
\begin{equation}
\Delta R \equiv \sqrt{ (\eta^{jet} -\eta^{b})^{2}
                      +(\phi^{jet} -\phi^{b})^{2} }.
\end{equation}
Of course, there is some ambiguity in this procedure,
but we have checked that more than 95\% of the $b$-jets
from the $h \to b\bar{b} $ decays are matched correctly.
Our procedure then allows us to apply a simplified parameterization
of the $b$-tagging efficiency to the same event sample simply rejecting
or accepting a given jet in the event with a certain probability.
 
\subsection{Backgrounds and optimal kinematical cuts}
 
Relevant backgrounds can be separated into two groups.
The first one contains irreducible background processes
with a $Z$ and two $b$-jets in the final state.
The second one contains all other potential backgrounds without a $Z$
or two $b$-jets in the final state.
In our analysis, we have applied double $b$-tagging to extract the signal
since after a preliminary analysis it quickly became evident
that single $b$-tagging is not sufficient.
 
Fig. 4 shows the invariant mass distribution for various backgrounds
$(d\sigma/d M_{llb\bar{b}})$ with a $b$-tagging efficiency of 40\%
and a purity of 2\% after the minimal cuts (I).
The dominant background is the irreducible process $pp \to Z b\bar{b} +X$,
provided that the $b$-tagging quality is high enough.
We estimate this process contribution using the $gg \to Z b\bar{b}$ subprocess
incorporated in PYTHIA \cite{Zbb}.
The total production cross section is about 600 pb at the LHC
and kinematics are in fairly good agreement with
the results of our parton level calculations.
To estimate the background from $pp \to Zjj + X, j={\rm jets}$,
when one or both non $b$-jets are mistagged, we use PYTHIA
to generate the $pp \to Zj +X$ processes with $p_T^j \ge 10$ GeV.
The second jet is generated in the parton shower evolution.
To avoid the double counting, events with two tagged $b$-jets
in the final state were eliminated from the $Zj$ sample.
The background from $t\bar{t} \to l\bar{l} \nu\bar{\nu} b\bar{b}$
is at the same level of $Z +$jets after the minimal cuts
and can be reduced further by a factor of 3 with the additional cut (II)
on the missing $E_T$,
\begin{equation}
E_T^{miss} \le 40 \; {\rm GeV}.
\label{eq:CUT2}
\end{equation}
The potentially dangerous background from $ZZ \to l\bar{l} b\bar{b}$
is below the level of the dominant $Z b\bar{b}$ by a factor of 10-20,
so the peak at $M_Z$ in the $M_{b\bar{b}}$ background distribution 
is nearly invisible.
 
We also estimated the backgrounds from $Z c\bar{c}$,
$c\bar{c} b\bar{b}$ and $b\bar{b} b\bar{b}$.
They are well below the uncertainty in our estimations of 
the $Zb\bar{b}$ background.
The additional isolation cut (III)
\begin{equation}
\Delta R (l,jet) \ge 0.3,
\end{equation}
effectively reduces the backgrounds with the leptons originating 
from $c$ or $b$ hadrons.
The $c$-jets are assumed to be mistagged as the $b$-jets with 
a probability of 9$\%$. 
 
We have unsuccessfully attempted to improve the significance of the signal
using harder kinematical cuts. 
It seems that to optimize significance
one has to use as soft cuts on $p_T^{l} $ and $ E_T^{jet} $
as possible taking into account detector limitations.
This conclusion is valid over almost the whole mass and $\tan\beta$ range
except some small region for $M_{A} > 330$ GeV and $\tan\beta > 1.5$,
where harder cuts provide some small increase in the significance
compared to the softest cuts used.
 
The last comment concerning Fig. 4 is that all background distributions
are peaked around  $M_{llbb} \simeq 200 $ GeV.
The large background makes the observation for the $A$
very difficult in this mass region
in spite of the sizeable production cross section of $pp \to A +X$
and a significant branching fraction of $A \to Zh$.
 
This discovery channel, $A \to Zh \to l\bar{l} b\bar{b}$, 
is especially attractive because it allows one, in principle,
to reconstruct the masses for both the $A$ and the $h$.
The optimal experimental procedure to extract the signal 
from the background seems to be an iterative one. 
At the first step, all $l\bar{l}jj$ events passing the cuts I,II,III,
and with both  $b$-tagged jets have to be preselected.
Then, in the case of known $M_h$, one can directly use the restriction
$|M_{jj} -M_{h}| \le \Delta M_{jj} $ to improve the observability
of the $A$ in the $ M_{lljj} $ distribution.
The window size $\Delta M_{jj}$ depends on the detector resolution.
For the CMS detector we estimate $\Delta M_{jj} \simeq 12 $ GeV
for this particular kinematical region.
It has to be stressed that the reconstructed $M_{b\bar{b}}$
has the systematic downward shift about 5-7  GeV
when one uses the usual cone jet-finding algorithms.
In the case that $M_h$ is beyond the reach of LEP II,
one can use some `sliding window' procedure.
Namely, one should scan the relevant 
$M_{jj} $ region by small steps, every time placing the middle of the 
window into the new position, trying to observe some bump in the 
$ M_{lljj }$ distribution. 
If there is some peak in the $M_{lljj}$ distribution,
one should apply the last step to improve observability for the $h$
(if its mass is unknown) by histogramming the $M_{jj}$ distribution
only for events with $M_{lljj}$ in some window around the $M_{A}$.
 
In Fig. 5, we illustrate the procedure of extracting the signal
for $M_{A} = 250, 345$ GeV and $\tan \beta = 1.5$.
A signals in the first row are obtained using $M_{jj}$ mass window shifted
with respect to the $M_{h}$ position.
The second row presents the $M_{llbb}$ distributions provided the optimal
$M_{jj}$ position is chosen.
And the last row demonstrates the $h$ signal over the background provided
the events are selected within the proper $M_{llbb}$ mass range
around the bumps found.
 
\subsection{Results}
 
We present the $5\sigma$ significance contours for the CMS detector with 
an integrated luminosity of $L_{int} =$ 10 and 30 fb$^{-1}$ in Fig. 6.
All the cuts I,II,III, have been applied and the sliding window procedure 
of Fig. 5 has been used to enhance the signal. 
We show results for two sets of the MSSM parameters: 
(1) $m_{\tilde{g}} \simeq m_{\tilde{l}} \simeq m_{\tilde{q}} =  \mu = 1$ TeV 
and 
(2) $m_{\tilde{g}} \simeq m_{\tilde{l}} \simeq m_{\tilde{q}} = -\mu = 500$ GeV.
 
We assume here that
$ L_{int} = 30$ fb$^{-1}$ will be accumulated 
at the low luminosity regime ($ \simeq 10^{33} cm^{-2} $) 
and that pile up effects will be small.
A higher luminosity might introduce 
additional complications due to the pile-up effects.
However, if the reasonable $b$-tagging efficiency (40\% + 2\%) 
could be achieved at a higher luminosity (about $10^{34} cm^{-2})$, 
the region explorable with $L_{int} = 30$ fb$^{-1}$ 
and a better $b$-tagging efficiency 60\% + 1\% 
can also be probed with the more modest $b$-tagging efficiency (40\% + 2\%) 
at $L_{int} = 100$ fb$^{-1}$.

For $m_{\tilde{g}} \simeq m_{\tilde{l}} \simeq m_{\tilde{q}} = \mu = 1$ TeV,
the signal of $A \to Zh \to l^+l^-b\bar{b}$ 
can be observable for $\tan\beta < 3$ and 180 GeV $< M_A \le 2m_t$.
For $m_{\tilde{g}} \simeq m_{\tilde{l}} \simeq m_{\tilde{q}} = -\mu = 500$ GeV,
this discovery channel, $A \to Zh \to l^+l^-b\bar{b}$, 
can be observable for $\tan\beta < 1.8$ and 180 GeV $< M_A \le 2m_t$.
 
\section{Conclusions}

The CP odd state of the electroweak symmetry breaking sector of
the MSSM is very difficult to observe at the LHC. 
We have found that $pp \to A \to Zh \to l^+l^-b\bar{b} +X$ 
may provide a promising opportunity to
simultaneously search for the $A$ and the $h$ at the LHC.
If SUSY particles are too heavy to be produced via decays of $A$, 
the signal of $A \to Zh \to l^+l^-b\bar{b}$ can be observable 
for $M_Z +M_h < M_A \le 2m_t$ and $\tan\beta < 3$.

The impact of SUSY decay modes on the $A \to Zh$ decay is significant.  
\begin{enumerate}
\item 
For $m_{\tilde{g}}\simeq m_{\tilde{l}}\simeq m_{\tilde{q}} = -\mu = 500$ GeV, 
the signal of $A \to Zh \to l^+l^-b\bar{b}$ can be observable 
for $\tan\beta < 1.8$. 
\item For a positive $\mu$ 
with $m_{\tilde{g}} \simeq m_{\tilde{l}} \simeq m_{\tilde{q}} = \mu$, 
the signal of $A \to Zh \to l^+l^-b\bar{b}$ might be observable 
only if $\mu > 600$ GeV and $\tan\beta$ is close to one.
\item 
For $m_{\tilde{g}}\simeq m_{\tilde{l}}\simeq m_{\tilde{q}} = |\mu| = 300$ GeV, 
SUSY decay modes reduce the signal to below the observable level. 
In this case, new channels to search for the $A$ become available. 
The most promising discovery mode is the $4l$ signal from the decay 
$A \to \tilde{Z}_2 \tilde{Z}_2 \to 4l +\tilde{Z}_1 \tilde{Z}_1$ \cite{Z2Z2}.
\end{enumerate}

\acknowledgements

This research was supported in part
by the U.~S. Department of Energy grants DE-FG-05-87ER40319, DE-FG02-91ER40685,
and DE-FG-03-94ER40833.



%


\begin{figure}
\caption[]{
The total cross section for the process 
$pp \to A \to Zh \to l^+l^- b\bar{b} +X$ in fb, 
as a function of $M_A$, for $\sqrt{s} = 14$ TeV, $m_t = 175$ GeV, 
$m_{\tilde{g}} \simeq m_{\tilde{l}} \simeq m_{\tilde{q}}$, 
and $\tan \beta$ = 1, 3, 10, and 30. 
The soft SUSY breaking parameters $A_t$ and $A_b$ are taken to be zero.
We take $m_{\tilde{g}} \simeq m_{\tilde{l}} \simeq m_{\tilde{q}} = |\mu|$  
and consider three cases: 
(a) $ \mu =$ 1000 GeV, 
(b) $ \mu =$  300 GeV, and 
(c) $-\mu =$  300 GeV. }
\end{figure}

\begin{figure}
\caption[]{
The total cross section of $pp \to A \to Zh \to l^+l^- b\bar{b} +X$ in fb,
as a function of $\tan\beta$, for $\sqrt{s} = 14$ TeV, $m_t = 175$ GeV,
and $M_A$ = 100, 200, 300, and 400 GeV.
Three cases are considered: 
(a) $m_{\tilde{q}} = \mu =$ 1000 GeV, 
(b) $m_{\tilde{q}} = \mu =$  300 GeV, and 
(c) $m_{\tilde{q}} =-\mu =$  300 GeV. 
Other parameters are the same as in FIG. 1.}
\end{figure}

\begin{figure}
\caption[]{
Contours in the $(M_A,\tan\beta)$ plane,
for $\sigma(pp \to A \to Zh \to l^+l^- b\bar{b} +X) =$ 1, 10, and 100 fb.
Three cases are considered: 
(a) $m_{\tilde{q}} = \mu =$ 1000 GeV, 
(b) $m_{\tilde{q}} = \mu =$  300 GeV, and 
(c) $m_{\tilde{q}} =-\mu =$  300 GeV. 
Other parameters are the same as in FIG. 1.}
\end{figure}

\begin{figure}
\caption[]{
Invariant mass distributions ($d\sigma/dM_{l\bar{l}b\bar{b}}$)
for the dominant SM backgrounds after cuts from:
$pp \to Z+2$ (non-b) jets $(Zj)$;
$q\bar{q} \to ZZ \to Zb\bar{b}$ $(ZZ)$;
$gg \to Zb\bar{b}$ $(Z b\bar{b})$ and
$gg \to t\bar{t}$ $(t\bar{t})$.
Double $b$-tagging
with 40$\%$ of $b$-tagging efficiency and 2$\%$ of mistagging is
assumed.
This figure is generated from a simulation with the CMS performance.}
\end{figure}
 
\begin{figure}
\caption[]{
The illustration for the sliding window procedure 
to enhance the signal to background ratio 
as described in Sec. III(C) of the text. 
This figure is generated from a simulation with the CMS performance,
for $m_{\tilde{g}}\simeq m_{\tilde{l}}\simeq m_{\tilde{q}} = \mu =$ 1000 GeV,
$\tan\beta = 1.5$ and $\sqrt{s} = 14$ TeV.
The left three figures are for $M_{A} = 250$ GeV, 
and the right ones for $M_{A} = 345$ GeV. 
The arrows in the second row of figures mark the chosen $M_{llbb}$ windows.}
\end{figure}
 
\begin{figure}
\caption[]{
The 5 $\sigma$ contour in the $M_A$ versus $\tan\beta$ plane, 
for $pp \to A \to Zh \to l^+l^- b\bar{b} +X$ in the MSSM, at the LHC.
This figure is generated using a simulation with the CMS performance, 
for $m_{\tilde{g}} \simeq m_{\tilde{l}} \simeq m_{\tilde{q}} = |\mu|$ 
with $\mu = 1000$ GeV (solid) and $\mu = -500$ GeV (dashed).
The figure on the left corresponds to an integrated luminosity 
of 10 fb$^{-1}$, the one on the right to that of 30 fb$^{-1}$. 
Case a) is for 60$\%$ of $b$-tagging efficiency and 1$\%$ mistagging, 
case b) is for a tagging (mistagging) efficiency of 40$\%$ (2$\%$).}
\end{figure}

\end{document}